\documentclass[aps,twocolumn,groupedaddress]{revtex4}
\usepackage[dvips]{color}
\usepackage{graphicx,color}
\begin{document}

\title{Majorana Flat Bands and Uni-directional  Majorana Edge States in Gapless Topological Superconductors}

\author{Chris L. M. Wong$^1$, Jie Liu$^1$, K. T. Law$^1$ and Patrick A. Lee$^2$}

\affiliation{1. Department of Physics, Hong Kong University of Science and Technology, Clear Water Bay, Hong Kong, China \\
2. Department of Physics, Massachusetts Institute of Technology, Cambridge, MA, USA}

\begin{abstract}
In this work, we show that an in-plane magnetic field can drive a fully gapped $p \pm i p$ topological superconductor into a gapless phase which supports symmetry protected Majorana edge states (MESs). Specifically, an in-plane magnetic field can close the bulk gap and create zero energy Majorana flat bands (MFBs) in the excitation spectrum. We show that the MFBs in the gapless regime are protected by a chiral symmetry and are associated with MESs.  Novel uni-directional MESs which propagate in the same direction on opposite edges appear when the chiral symmetry is broken. The MFBs and the uni-directional MESs induce nearly quantized zero bias conductance in tunneling experiments which are robust in the presence of a gapless bulk and disorder.
\end{abstract}


\maketitle

\emph{Introduction}---A topological superconductor (TS) has a bulk superconducting gap and topologically protected gapless boundary states [\onlinecite{SRFL, Kitaev, Teo, Qi, Beenakker1}]. TSs are under intense theoretical and experimental studies due to the possibility of realizing Majorana fermions in these systems, which act as their own antiparticles and obey non-Abelian statistics [\onlinecite{RG, Kitaev00, Ivanov}]. Majorana fermions in TSs are topologically protected, in the sense that the Majorana fermions cannot be removed by perturbations unless the bulk energy gap is closed or certain symmetries are broken. 

Remarkably, recent development shows that topologically protected Majorana fermions exist in systems where the bulk is gapless [\onlinecite{Beri, Sato1, Sato2, Sato3, Schnyder1, Schnyder2, Schnyder3, Sato4, Sau, Balents, WangFa}]. For example, Majorana edge states (MESs) with flat dispersion can be found in 2D nodal $d_{xy}+p$-wave superconductors which respect time-reversal symmetry [\onlinecite{Sato1, Sato2, Sato3}]. It is also shown that zero energy Majorana flat bands (MFBs) can appear on the surface of 3D time-reversal invariant non-centrosymmetric superconductors which have topologically stable line nodes in the bulk [\onlinecite{ Schnyder1, Schnyder2, Schnyder3}]. Proposals on gapless TSs which break time-reversal symmetry have been made recently [\onlinecite{ Sato4, Sau, Balents}].

All the cases mentioned above are intrinsic gapless superconductors. In this work, we show that an in-plane magnetic field can drive a fully gapped  $p \pm ip$-wave TS into a gapless regime which supports symmetry protected Majorana fermions. An in-plane magnetic field may first close the bulk gap. Further increasing the strength of the magnetic field creates zero energy MFBs in the excitation spectrum when the bulk is gapless. Importantly, we show that the MFBs are protected by a chiral symmetry and are robust against disorder. The evolution of the excitation spectrum of a  $p \pm ip$-wave superconductor as a function of the in-plane magnetic field strength is shown in Fig.1 and Fig.2.

S-wave pairing and Rashba spin-orbit coupling terms break the chiral symmetry which protects the MFBs and lift the Majorana fermions from zero energy. In this case, uni-directional MESs, which are distinct from the usual helical or chiral MESs in that the modes on opposite edges move in the same direction, may appear. Finally, we show that the MESs survive in the presence of disorder and induce nearly quantized zero bias conductance in tunneling experiments.

\begin{figure}
\includegraphics[width=3in]{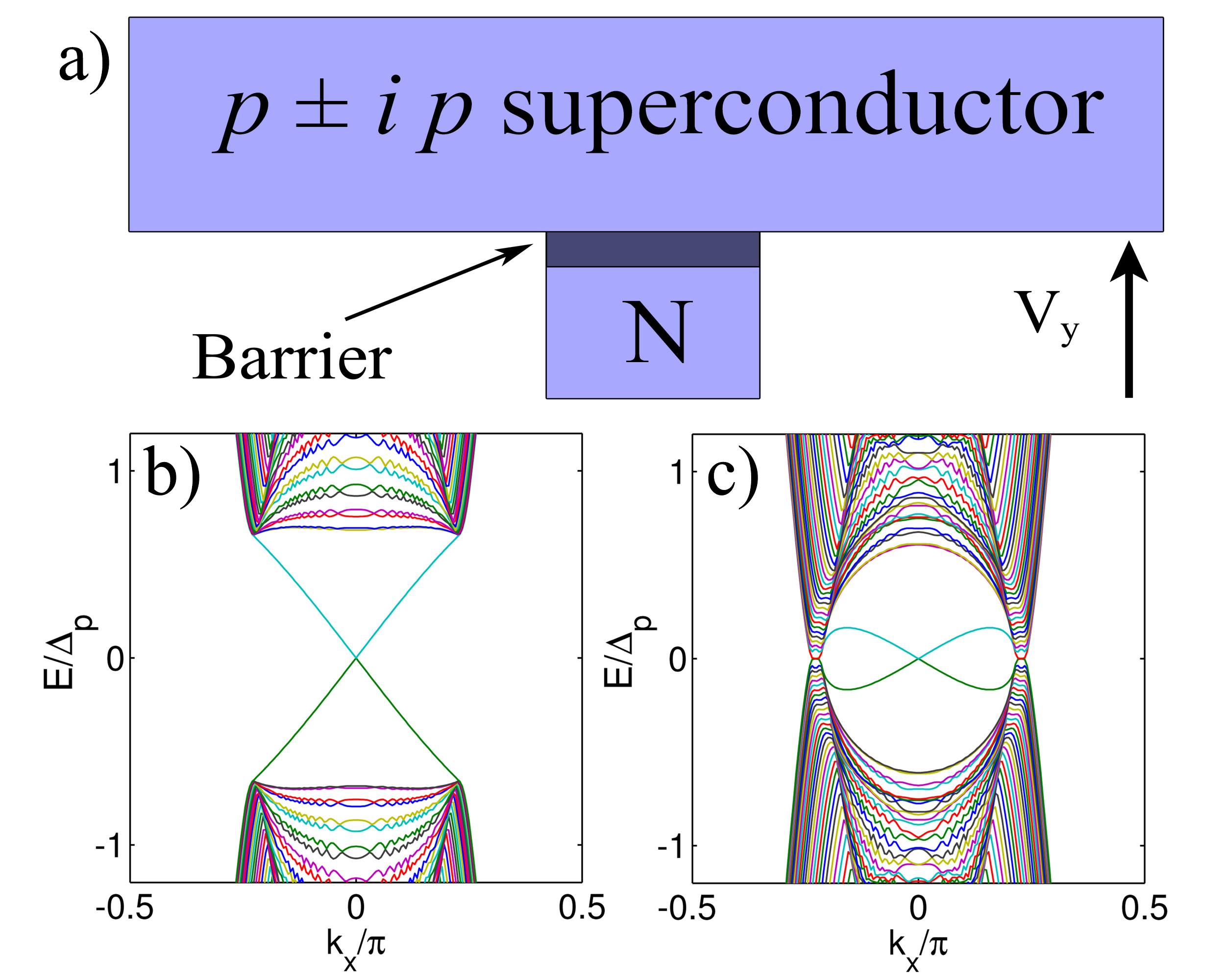}
\caption{\label{Fig1} a) A schematic picture of a $p \pm ip$-wave superconductor subject to an in-plane magnetic field $V_y$. A tunnel junction and a normal lead N is attached to the superconductor.   b) The energy spectrum of a $p \pm ip$ superconductor in the  topologically non-trivial regime. Periodic boundary conditions in the $x$-direction and open boundary conditions in the $y$-direction are assumed. The parameters are $t=12\Delta_p$, $\mu=3\Delta_p -2t$, $\Delta_s=0$, $\alpha_R=0$ and $V_y=0$. c) Same parameters as b), except $V_y=0.7\Delta_p$. The bulk energy gap is closed in this regime. }
\end{figure}

\emph{Majorana Flat Bands}--- We start with a BdG Hamiltonian which describes a $p \pm i p$-wave superconductor
\begin{equation}
H_{p}(\mathbf{k})= \left(
\begin{array}{cc}
 \xi(\mathbf{k}) + \mathbf{V} \cdot \mathbf{\sigma} & \hat{\Delta}(\mathbf{k}) \\
 \hat{\Delta}^\dagger (\mathbf{k}) &  -\xi^T (\mathbf{-k})  - \mathbf{V} \cdot \mathbf{\sigma^*}
\end{array} \right) .
\end{equation}

Here $\xi(\mathbf{k}) =[-t (\cos k_{x} + \cos k_{y})- \mu] \sigma_0 - \alpha_R[-\sin k_y\sigma_x + \sin k_x \sigma_y]$ is the sum of the kinetic energy and the Rashba spin-orbit coupling, $\mathbf{V}$ describes the Zeeman coupling of the electrons with an external magnetic field, $\hat{\Delta}(\mathbf{k}) = (\Delta_s + \mathbf{d(k) \cdot \sigma}) (i \sigma_{y})$  is the superconducting gap function. We first assume that the spin-singlet pairing amplitude $\Delta_s$ and the Rashba spin-orbit coupling $\alpha_R$  are zero. The spin-triplet pairing vector is chosen as  $\mathbf{d(k)}= \Delta_p (-\sin k_{y}, \sin k_{x}, 0)$ such that the Hamiltonian descirbes a two dimensional $p \pm i p$-wave superconductor where $\Delta_p$ is a constant. When $\mathbf{V}=0$, the Hamiltonian respects both time-reversal symmetry $T= U_{T}K$ with $U_{T}^{-1} H_{p}^{*}(\mathbf{k}) U_{T} =  H_{p}(\mathbf{-k})$ and particle-hole symmetry $P= U_{P}K$ with $U_{P}^{-1} H_{p}^{*}(\mathbf{k}) U_{P} =  -H_{p}(\mathbf{-k})$. Here, K is the complex conjugate operator, $ U_{T}=\sigma_0 \otimes i\sigma_y$ and $U_{P}=\sigma_x \otimes \sigma_0 $ such that $T^2=-1$ and $P^2=1$.

According to symmetry classification, the above Hamiltonian in the absence of an external magnetic field belongs to DIII class which can be topologically non-trivial [\onlinecite{SRFL}]. In the topologically non-trivial regime, the $p \pm ip $ superconductor possesses gapless counter propagating helical MESs. The energy spectrum in the topologically non-trivial regime is shown in Fig.1b. In the rest of this section, we show that the $p \pm ip$ superconductor responds to an in-plane magnetic field in an anomalous way as described in the \emph{Introduction}.

\begin{figure}
\includegraphics[width=3.2in]{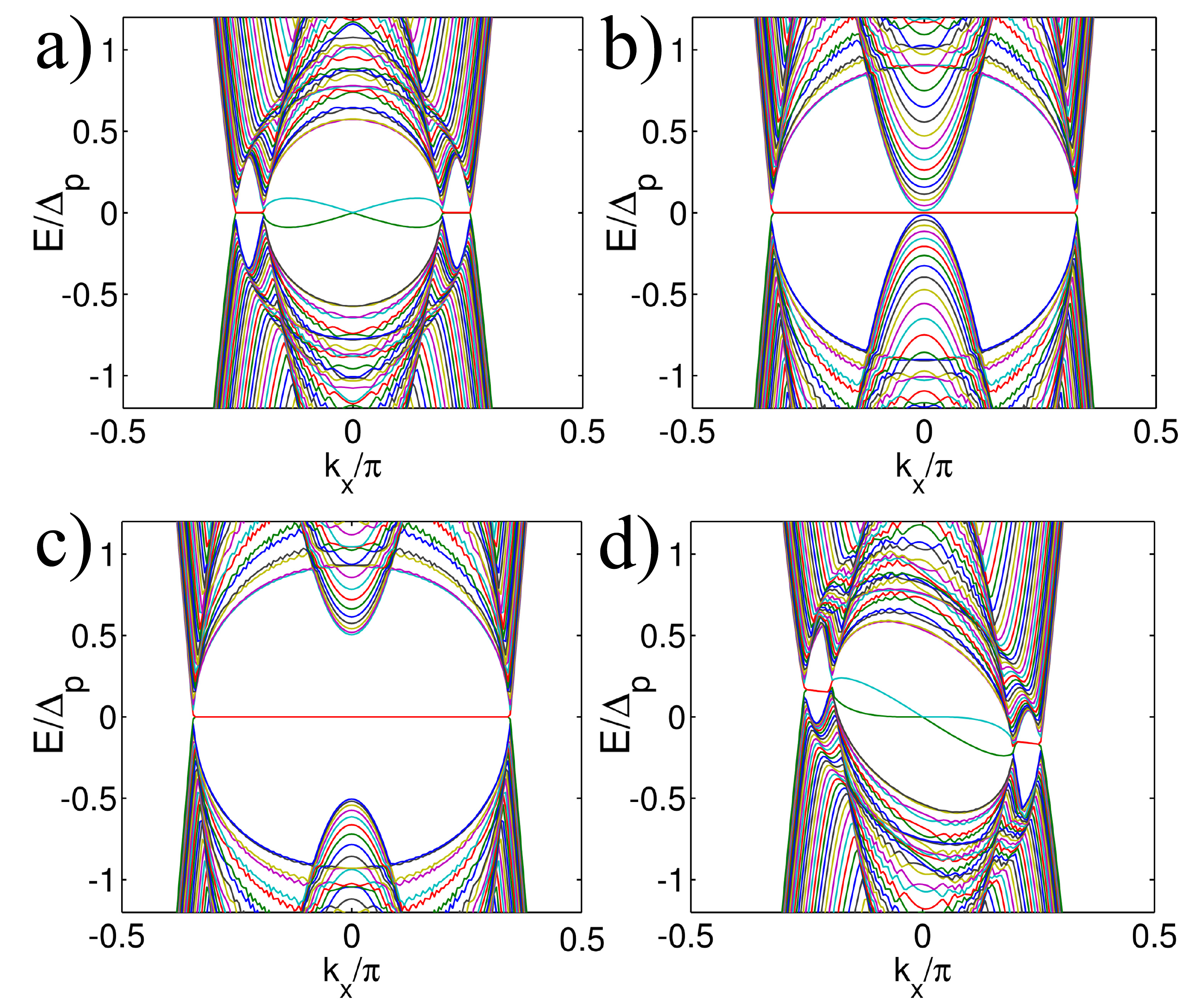}
\caption{\label{Fig2} The evolution of the energy spectrum of a $p \pm i p$-wave superconductor as $V_y$ increases. For a) to c) the parameters are the same as Fig.1b except the values of $V_y$.  a)  $V_y=\Delta_p$. b) $V_y=3\Delta_p$. c) $V_y=3.5\Delta_p$. d) S-wave pairing and Rashba terms with $\Delta_s=0.3\Delta_p$ and $\alpha_R=0.2 \Delta_p$ are added to a).}
\end{figure}

To be specific, we suppose a magnetic field is applied in the y-direction such that $\mathbf{V}=(0,V_y,0)$. In the presence of a magnetic field, the time-reversal symmetry $T=U_{T}K$ is broken. However, one can show that the Hamiltonian satisfies a time-reversal like symmetry $T_{1d}=U_{T1d}K$ such that $T_{1d}^{-1} H(k_x,k_y) T_{1d}= H(k_x, -k_y)$, where $U_{T1d}= \sigma_{z} \otimes \sigma_{z}$. Moreover, the Hamiltonian satisfies a particle-hole like symmetry $P_{1d}= U_{P1d}K $ such that $ P^{-1}_{1d} H(k_x,k_y) P_{1d} = - H(k_x, -k_y)$ with $U_{P1d}= \sigma_y \otimes \sigma_y$. Due to the fact that the symmetry operators operate on $k_y$ only and $k_x$ is unchanged, one may regard $k_x$ as a tuning parameter and the Hamiltonian can be written as $H_{k_x}(k_y)$. Since $H_{k_x}(k_y)$ satifies the symmetries $T_{1d}$ and $P_{1d}$ with $T_{1d}^2=P_{1d}^2=1$, $H_{k_x}(k_y)$ is a BDI class Hamiltonian which can be classified by an integer [\onlinecite{SRFL}].

To classify the Hamiltonian $H_{k_x}(k_y)$ with $k_x$ as a tuning parameter, we note that as a result of the $T_{1d}$ and $P_{1d}$ symmetries, $H_{k_x}(k_y)$ satisfies the chiral symmetry $S_{1d}=T_{1d}P_{1d}$ with 
\begin{equation}
S_{1d}^{-1} H(k_x,k_y)S_{1d}= -H(k_x,k_y).   \label{chiral}
\end{equation}
In this case, $H_{k_x}(k_y)$ can be off-diagonalized in the basis which diagonalizes $S_{1d}$ such that
\begin{equation}
\tilde{H}_{k_x}(k_y) = \left(
\begin{array}{cc}
 0 & A_{k_x}(k_y) \\
A_{k_x}^{\dagger}(k_y) & 0
\end{array} \right),
\end{equation}
Note that $A_{k_x}(k_y)$ is real at $k_y=0, \pm \pi$, we can define the quantity
\begin{equation}
z(k)=e^{i \theta(k)}=\text{Det}[A_{k_x}(k)]/|\text{Det}[A_{k_x}(k)]|,
\end{equation}
such that $\theta(k) = n \pi$ at $k=0, \pm \pi$ with integer $n$. The winding number of $\theta(k)$ can be used as the topological invariant which characterizes the Hamiltonian $H_{k_x}(k_y)$. The winding number $N_{BDI}$ can be written as \cite{Tewari}
\begin{equation}
N_{BDI}=\frac{-i}{\pi} \int_{k_y=0}^{k_y=\pi} \frac{dz(k_y)}{z(k_y)}.   \label{winding}
\end{equation}
Using $A_{k_x}(k_y)$ obtained from $H_{k_x}(k_y)$, it can be shown that $|N_{BDI}|=1$ when
\begin{equation}
\begin{array}{l}   \label{NBDI}
\mathcal{M}(k_x,k_y=0)  \mathcal{M}(k_x, k_y=\pi) <0,  \quad \text{where} \\
\mathcal{M}(k_x,k_y)= [\mu + t (\cos k_x+ \cos k_y)]^2 + \Delta_p^2 \sin^2 k_x  - V^2_y,
\end{array}   
\end{equation}
assuming that $V_y$ and $\Delta_p$ are non-zero. In the range of $k_x$ where $N_{BDI}=1$, the Hamiltonian $H_{k_x}(k_y)$ is topologically non-trivial. For a $p \pm ip$ superconductor with periodic boundary conditions in the $x$-direction and open boundary conditions in the $y$-direction, there are zero energy Majorana modes localized on the edges of the system when Eq.\ref{NBDI} is satisfied. Therefore, MFBs appear in the corresponding parameter regime. 

The evolution of the energy spectrum of a $p \pm ip $ superconductor as a result of an increasing in-plane magnetic field is shown in Fig.1 and Fig.2. First, an in-plane magnetic field reduces the bulk gap as shown in Fig.1c. Second, after the bulk gap is closed, MFBs appear for a finite range of $k_x$ where $|N_{BDI}|=1$ as shown in Fig.2a. Third, by further increasing the magnetic field, the bulk gap at $k_x=0$ is closed (Fig.2b). Fourth, by increasing the magnetic field even further,  the energy crossing at $k_x=0$ disappears and only a MFB remains (Fig.2c). It is important to note that the MFBs appear when the bulk is gapless. The bulk energy spectrum of a $p \pm ip$-wave superconductor corresponding to Fig.2a is shown in Fig.3a. It is evident that there are nodal points in the bulk spectrum when MFBs appear. The nodal points in Fig.2a are the projection of the bulk nodal points on the $k_x$-axis in Fig.3a, similar to the cases in intrinsic gapless TSs [\onlinecite{Schnyder3, Sato3, WangFa}]. Both the nodal points in the bulk spectrum as well as the MFBs are protected by the topological invariant $N_{BDI}$. In other words, the MFBs and the nodal points in the bulk appear whenever $N_{BDI}$ is non-trivial for some range of $k_x$.

\begin{figure}
\includegraphics[width=3.2in]{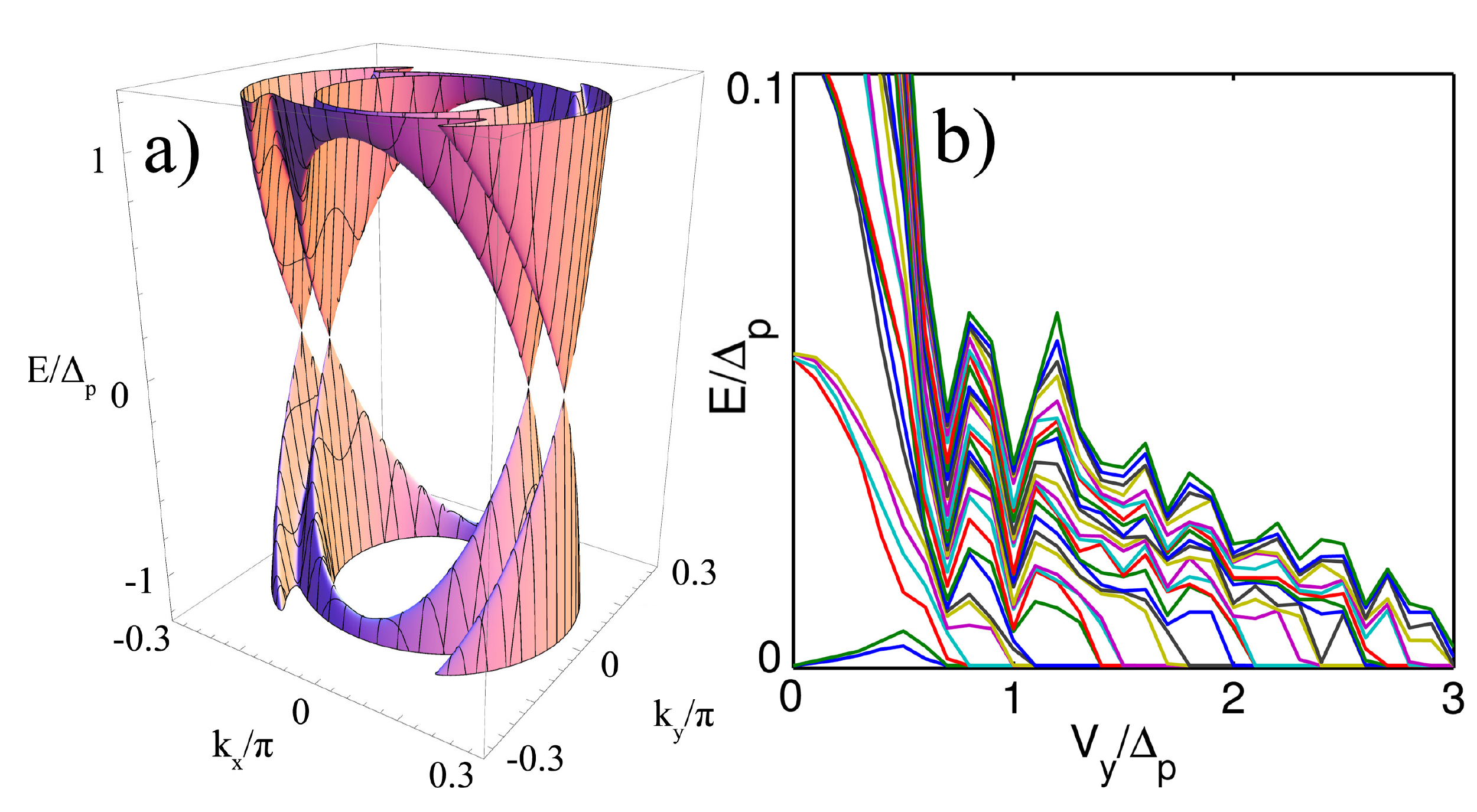}
\caption{\label{Fig3} a) The bulk energy spectrum of a $p \pm ip$ superconductor. The parameters are the same as the ones in Fig.2a but periodic boundary conditions in both the $x$ and $y$ directions are imposed. b) The energy spectrum of a pure $p \pm ip$-wave superconductor with dimensions $L_x=100a$ and $L_y=300a$,where $a$ is the lattice spacing. Only the thirty lowest energy eigenvalues are shown.  Periodic boundary conditions in the $x$-direction and open boundary conditions in the $y$-direction are assumed. On-site Gaussian disorder with variance $\text{w}^2=(1.5\Delta_p)^2$ is present. It is evident that as Vy increases, states collapse to zero energy and stay there, increasing the number of zero energy modes. This indicates the widening of the MFBs as $V_y$ increases. Importantly, the zero energy Majorana modes are not lifted by disorder.  }
\end{figure}

\emph{Uni-directional Majorana Edge States}--- It is shown above that MFBs appear when $N_{BDI}=1$ for a finite range of $k_x$ and the MFBs are protected by the chiral symmetry $S_{1d}$ in Eq.\ref{chiral}. However, s-wave pairing and Rashba terms, which can arise in non-centrosymmetric superconductors [\onlinecite{Tanaka}], break the chiral symmetry and lift the zero energy modes to finite energy as shown in Fig.2d. In the case of  adding s-wave and Rashba terms to Fig.2c, the MFB acquires a finite slope and uni-directional MESs appear at the sample edge as shown in Fig.4a. A schematic picture of the uni-directional MESs is shown in the insert.

Another interesting finding is that the uni-directional MESs can appear in the absence of $p \pm ip$-wave pairing. The energy spectrum of an s-wave superconductor with Rashba terms and finite $V_y$ is shown in Fig.4b. It is evident that uni-directional MESs appear in this case. To understand the origin of the MESs in the gapless phase, we note that the Hamiltonian $H_{p}(k_x,k_y)$ at $k_x=0$ satisfies the chiral symmetry $S=PT_{1d}$  and is classified by the topological invariant $N_{BDI}$ when $\Delta_p=0$. If both the $\Delta_s$ and $\alpha_R$ are non-zero, the Hamiltonian at $k_x=0$ has winding number $|N_{BDI}|=1$ when
\begin{equation}
\begin{array}{l}    \label{swaveN}
\mathcal{M}_s (0,0) \mathcal{M}_s (0,\pi) <0,    \quad \text{where} \\
\mathcal{M}_s(k_x,k_y)=[\mu + t (\cos k_x + \cos k_y)]^2+\Delta_s^2 - V^2_y.
\end{array}
\end{equation}
Non-trivial $N_{BDI}$ indicates the appearance of the uni-directional MESs. It is important to note that systems with pure s-wave pairing and Rashba terms can be realized by inducing s-wave superconductivity in semi-conductors as demonstrated in recent experiments [\onlinecite{Mourik, Deng, Das}]. This opens a way for realizing the novel uni-directional MESs. 

\begin{figure}
\includegraphics[width=3.2in]{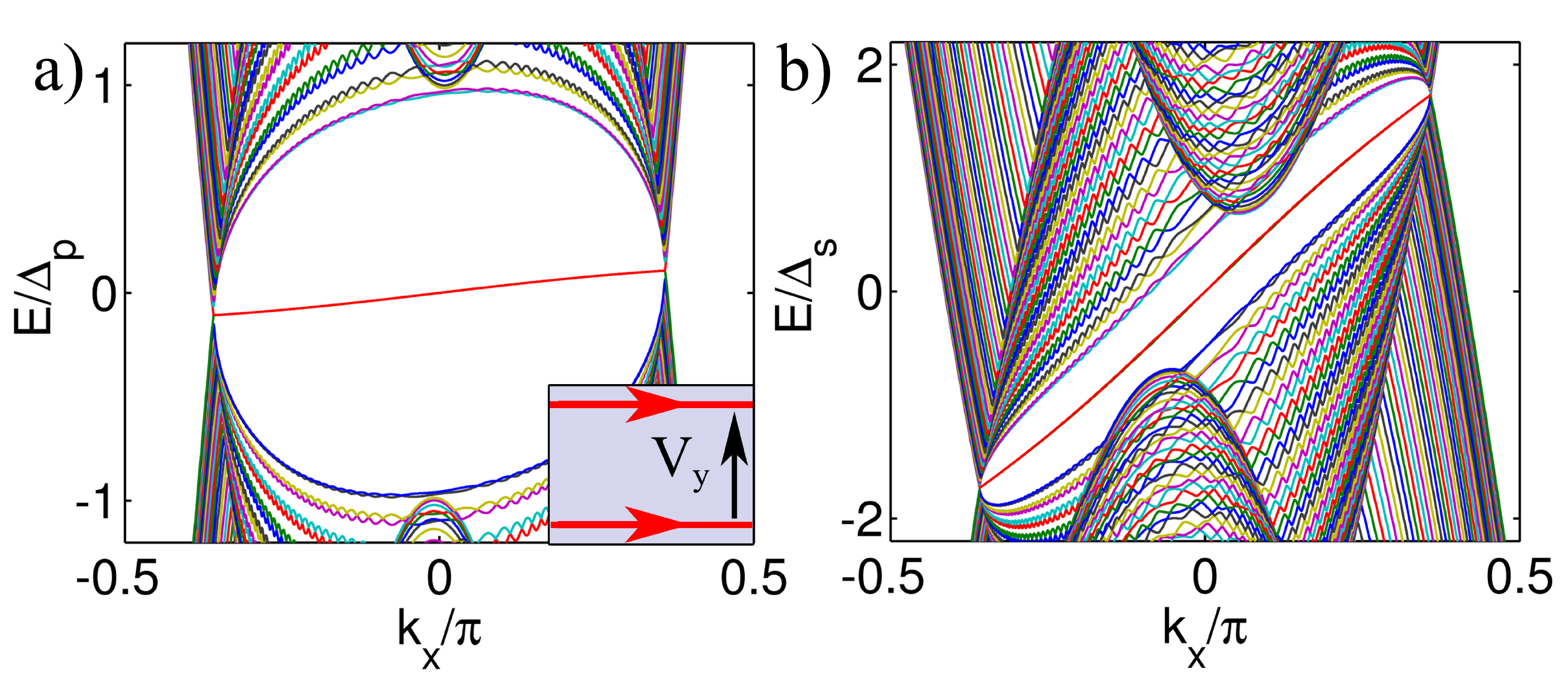}
\caption{\label{Fig4} a) The parameters are the same as those in Fig.2d except the values of $V_y$. The MFB acquires a finite slope when $\Delta_s$ and $\alpha_R$ are finite at $V_y=4\Delta_p $. Uni-directional  MESs appear in this regime.  A schematic picture of the uni-directional MESs is shown in the insert. b) The energy spectrum in the presence of spin-singlet pairing and Rashba and no $p$-wave pairing terms. Here, $\Delta_p=0$, $t=12 \Delta_s $, $\mu=3\Delta_s -2t$ and $\alpha_R=2 \Delta_s$.}
\end{figure}

\emph{Experimental Detection and Effects of Disorder} --- It has been shown in previous works that Majorana fermions induce resonant Andreev reflection at the junction between a normal lead and a fully gapped TS [\onlinecite{LLN, WADB}]. However, resonant Andreev reflection may not happen when the bulk is gapless due to the non-vanishing direct tunneling amplitudes from the normal lead to the gapless superconductor. Finite direct tunneling amplitudes make the reflection matrix non-unitary and render the arguments leading to resonant Andreev reflection not applicable [\onlinecite{WADB}]. In this section, we calculate the zero bias conductance (ZBC) of a junction between a normal lead and a TS as a function of the in-plane magnetic field strength. It is found that MFBs and uni-directional MESs  induce nearly quantized ZBC even when the bulk is gapless and in the presence of disorder.

A schematic picture of the experimental setup is shown in Fig.1a. A normal lead is coupled to an edge of the TS to form a NS junction. Using the lattice Green's function method [\onlinecite{Sun, Lee, Fisher}], we calculate the direct tunneling amplitude and the Andreev reflection amplitude of the NS junction. The results for the ZBC, $\frac{dI}{dV}$ at zero voltage bias $V=0$, are depicted in Fig.5. 

\begin{figure}
\includegraphics[width=3.2in]{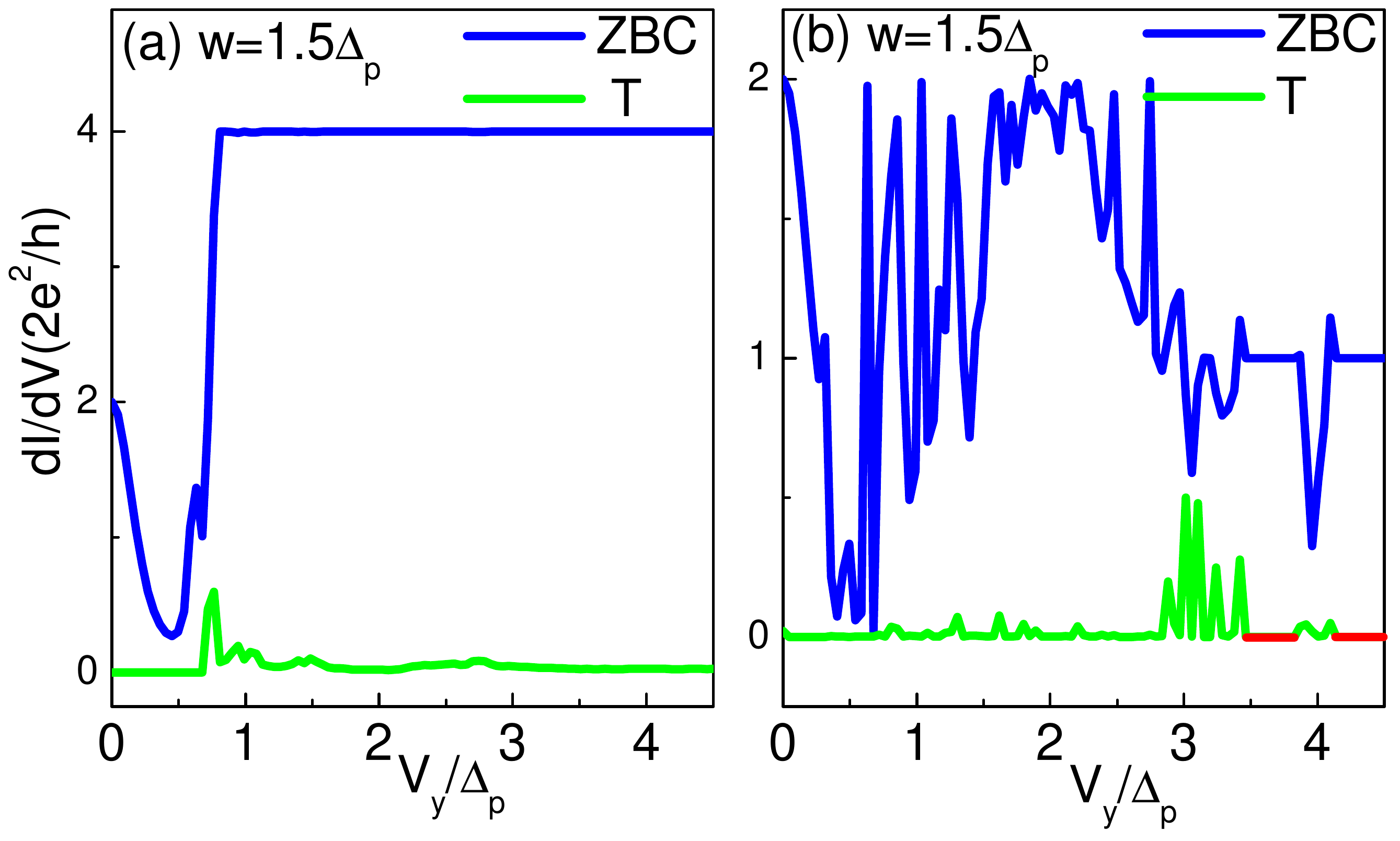}
\caption{\label{Fig5} The ZBC versus $V_y$. a) ZBC from a lead to a  $p \pm ip$-wave superconductor. The setup is depicted in Fig.1a. The superconductor has dimensions $L_x=100a$ and $L_y=300a$. Periodic boundary conditions in the x-direction is assumed. $t=12\Delta_p$, $\mu=3\Delta_p-2t$ in the superconductor and the lead. The barrier between the lead and the superconductor is simulated by a reduced hopping amplitude $t_c = 0.3t$. A semi-infinite lead with width $8a$ is used in the simulation. The number of conducting channels in the lead is $N_c =4$. The direct tunneling contribution to the ZBC is denoted as $T$.  Gaussian on-site disorder with variance $\text{w}^2=(1.5 \Delta_p)^2$ is present. b) S-wave pairing and Rashba terms with $\Delta_s =0.3 \Delta_p$ and $\alpha_R = 0.2\Delta_p$ are added to a). The red lines near $V_y =4 \Delta_p $ indicate the regime where the ZBC is quantized due to the presence of the uni-directional MESs. }
\end{figure}

Fig.5a shows the ZBC as a function of $V_y$ for a $p \pm ip$-wave superconductor in the presence of on-site disorder. The ZBC depends on $V_y$ and  the number of channels $N_c$ in the normal lead. The ZBC for a lead with $N_c=4$ (including spin degeneracy) is plotted in Fig.5a. The direct tunneling contribution to the ZBC, $T$, is also shown.

To understand the results, we note that time-reversal symmetry is preserved and the system is fully gapped at $V_y=0$,  the $4e^2/h$ quantization of ZBC is the property of a DIII class TS which has two Majorana zero modes on the edge [\onlinecite{Fulga, Wong}]. As $V_y$ increases, time-reversal symmetry is broken and the ZBC is suppressed by disorder. However, further increasing $V_y$ closes the bulk gap. When this happens, there is a large jump in the ZBC. This jump is due to the contribution from the Andreev reflection caused by the MFBs and the direct tunneling caused by the gapless bulk. This can be clearly seen from the $V_y$ dependence of $T$ in Fig.5a. It is interesting to note that the final ZBC is almost quantized at $\frac{2e^2}{h} N_c$ with $N_c=4$ due to the Andreev reflection caused by the large number of independent Majorana fermions from the flat band. In general, the ZBC caused by MFBs saturates at $\frac{2e^2}{h} N_c$.

The nearly quantized ZBC at large $V_y$ in Fig.5a suggest that the MFBs are robust against disorder. To confirm this,  the energy spectrum of the $p \pm ip$ superconductor with parameters corresponding to Fig.5a is shown in Fig.3b. It is evident from Fig.3b that finite energy states collapse to zero energy and stay there, increasing the number of zero energy modes as $V_y$ increases. 

Indeed, the MFBs are protected by the chiral symmetry $S_{1d}$ in Eq.\ref{chiral}. It can be shown that zero energy edge states, which are simultaneous eigenstates of the Hamiltonian and $S_{1d}$, have the same chirality for states localized on the same edge of the sample. The chirality of an eigenstate of $S_{1d}$ is defined as the eigenvalue of the state on $S_{1d}$, which is always $+1$ or $-1$. The net chirality number (NCN) of an edge of the sample, which is the sum of the chirality numbers of all the zero energy states localized on the edge, is always non-zero when MFB appears. Moreover, it can be shown that the number of stable zero energy modes on an edge equals the NCN of the edge [\onlinecite{Sato3}]. Since on-site disorder does not break the chiral symmetry and cannot change the NCN, the number of stable zero energy modes cannot be changed by disorder. 

It can be shown in the $p \pm ip$ superconductor case that opposite edges have opposite NCNs. Therefore, the zero energy modes can be removed only if the chiral symmetry $S_{1d}$ is broken or when two edge states with opposite chiralities are coupled to each other. A similar index theorem for time-reversal invariant TSs is first shown in Ref.[\onlinecite{Sato3}]. However, in the time-reversal invariant case, the NCN on an edge is always zero and the flat bands are not robust against disorder. 

Fig.5b shows the ZBC versus $V_y$ when s-wave pairing and Rashba terms are added to the Fig.5a. At $V_y=0$, the ZBC is quantized at $2\frac{2e^2}{h}$ as expected for a fully gapped DIII class TS [\onlinecite{Fulga, Wong}]. When $V_y$ is increased, the ZBC decreases due to disorder. Further increasing $V_y$ closes the bulk gap and there is a jump in the ZBC. Moreover, it is important to note that there are ZBC plateaus near $V_y=4\Delta_p$. As it is shown before, uni-directional MESs appear near this regime as shown in Fig.4a. Even though the bulk is gapless, the number of bulk states with zero energy is very small for certain parameter regimes. In this case, the direct tunneling amplitude is small and one can ignore direct tunnelings. As a result, the ZBC is quantized due to Majorana fermion induced resonant Andreev reflection [\onlinecite{LLN}]. 

\emph{Conclusion}--- We show that an in-plane magnetic field can drive a $p \pm i p$-wave superconductor to a gapless phase which supports MFBs. In the presence of s-wave pairing and Rashba terms, the MFBs acquire finite slopes and uni-directional MESs appear. These Majorana modes are symmetry protected and robust against disorder. They induce nearly quantized ZBC in  tunneling experiments.

\emph{Acknowledgments}--- The authors thank A. Akhmerov, C. Kane, T.K. Ng, Y. Tanaka and especially M. Sato for inspiring discussions.  CLMW, JL and KTL  are supported by HKRGC through DAG12SC01 and HKUST3/CRF09. KTL thanks the support of HKUST SSc Computational Science Initiative. PAL acknowledges the support from DOE Grant No. DEFG0203ER46076.

\end{document}